# What do we see when we look at networks
## an introduction to visual network analysis and force-directed layouts
*Tommaso Venturini, Mathieu Jacomy, Pablo Jensen*

It is an increasingly common practice in several natural and social sciences to rely on network visualisations both as heuristic tools to get a first overview of relational datasets and as a way to offer an illustration of network analysis findings. Such practice has been around long enough to prove that scholars find it useful to project networks on a space and to observe their visual appearance as a proxy for their topological features. Yet this practice remains largely based on intuition and no investigation has been carried out on to render explicit the foundations and limits of this type of exploration. This paper provides such analysis, by conceptually and mathematically deconstructing the functioning of force-directed layouts and by providing a step-by-step guidance on how to make networks readable and interpret their visual features.

## Introduction

There is no point in denying it, networks are not only mathematical but also visual objects. If graph mathematics is exploited in engineering and natural sciences since the 18th century, network visualisation has become increasingly popular in the last couple of decades. This visual renaissance is particularly noticeable in the digital humanities and social sciences, where the increasing availability of relational datasets has fuelled a renewed interest for graph charts, but it has also touched other disciplines such as ecology, neurosciences, and genetics. In general, it has become common in the scientific literature to illustrate social relations, economic fluxes, linguistic co-occurrences, proteins interactions, neuronal connections and many other relational phenomena as point and "points-and-lines charts".

The scientific function of such charts, however, is seldom questioned and often equivocal. While network visualisations are generally left out of the actual demonstration (which relies instead on calculations and metrics), they are regularly exhibited to provide a more tangible and intuitive grasp on the findings. We distrust network charts as scientific proofs and yet we embrace them for their insights.

This ambivalence, we believe, should not be denied, but rather explicitly addressed and used as a springboard to suggest a more mindful use of network visualisations. As we will argue, the very ambiguity that makes graph drawings unfit for hypothesis confirmation, makes them extremely useful for exploratory data analysis. This is particularly true for the kind of medium-sized networks often found in social and biological phenomena. Networks spanning from hundreds to thousands of nodes are too large to be charted manually, and yet too small to justify a rush for metrics and models. *Before* proceeding to aggregation and testing, *visual analysis* offers a tool to explore the richness of relational datasets that is increasingly widespread and yet surprisingly undocumented.

To address this lack of documentation, this paper offers a basic introduction to the technique of *visual network analysis* (VNA) and tries to make explicit its implicit foundations. Because this technique is yet unsettled, we will be obliged to alternate theoretical and practical considerations. (1) We start by reviewing the emerging standards of VNA and explain their conceptual and historical bases. (2) We provide practical advices on how to read graphs and make them more legible. (3) We discuss, more conceptually, the kind of information delivered by VNA and its specific way of dealing with ambiguity. (4) We conclude by sketching a formal analysis of limits of network visualisation and calling for the help of the mathematical community to better define such limits.

# 1. Force-directed layouts and the basis of visual network analysis

The heuristic value of network visualisations was first noticed in the second half of the 20th century in the early school of *social networks analysis* or SNA (Scott, 1991, Wasserman & Faust, 1994). Jacob Moreno, founder of such approach, explicitly affirmed that "the expression of an individual position can be better visualised through a sociogram than through a sociometric equation" (Moreno, 1934, p. 103).

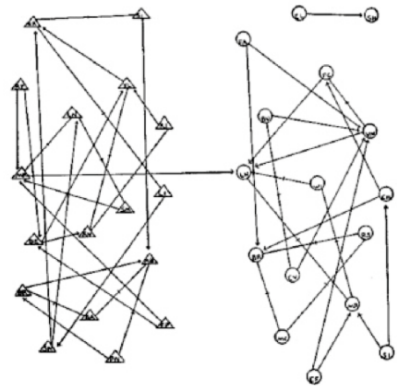

*Figure 1. Sociogram representing friendship among school pupils*
*(original title and image accompanying Jacob Moreno's interview by the New York Time, 1933)*

Working on their sociograms, Moreno and his disciples set the standards for the visualisation of networks (Freeman, 2000 & 2009). Their point-and-line approach to graph drawing has been so successful that it has become the *de facto* standard of network drawing. So much, that it feels useless to specify that in these charts the points represent the nodes and the lines represent the edges connecting them, although this choice is by no means evident (in matrices, for instance, points indicate connections while nodes are rendered as rows and columns).

But standardisation has gone further. Even within the *points-and-lines* family, diversity has been progressively reduced and today most networks visualisations abide by three unwritten principles according to which nodes are (1) positioned according with their connectivity; (2) sized proportionally to their importance; and (3) coloured or shaped by their category. Together these principles constitute the foundations of VNA. We will see, in the next section, how these principles combine to make networks legible. For the moment, let us consider the first one, which is the most specific to this technique but also the most problematic.

The cornerstone of VNA is the use of "force-directed layouts" to draw networks in a two-dimensional space (Battista *et al.*, 1999). These algorithms may be implemented according to different recipes but, ultimately, they all rest on the same physical analogy: nodes are charged with a *repulsive force* driving them apart, while edges introduce an *attractive force* between the nodes that they connect. Once launched, force-vectors vary the position of nodes trying to balance the repulsion of nodes and the attractions of edges. At equilibrium, force-directed layouts produce a disposition of nodes that is *visually meaningful*: nodes that have more direct or indirect neighbours *tend to* find themselves closer in the layout.

This technique for visualising graphs has become so common that we often fail to notice its remarkable accomplishment. Force-directed algorithms do not just project networks in space – they create a space that would not exist without them. This is why we name this process "spatialisation" rather than "visualisation". It does more than producing a convenient image, it embeds the network in a specific space in a process that has applications outside visualisation (where it is known as "graph embedding", Yan *et al.*, 2007). Spatialisation creates a space that retains key properties of a network.

To understand this feat, consider the plan of an underground, bus or train system. Most of these plans are not strictly geographical maps (i.e. drawn by setting a system of coordinates *first* and *then* placing the stations according to their coordinates), but charts in which proximity represent connectivity rather than bird-fly distance. Introduced in 1993 by Harry Beck for the London tube (Hadlaw, 2003), this design technique has become a standard for public transportation system. Compared to geographic maps, this type of representation is more focused on the information needed by users (how long it will take to go from A to B) while remaining readable according to the same visual conventions. Not a little advantage given the huge efforts invested to build and spread the conventions of cartographic mapping (see, among others, Turnbull, 2000).

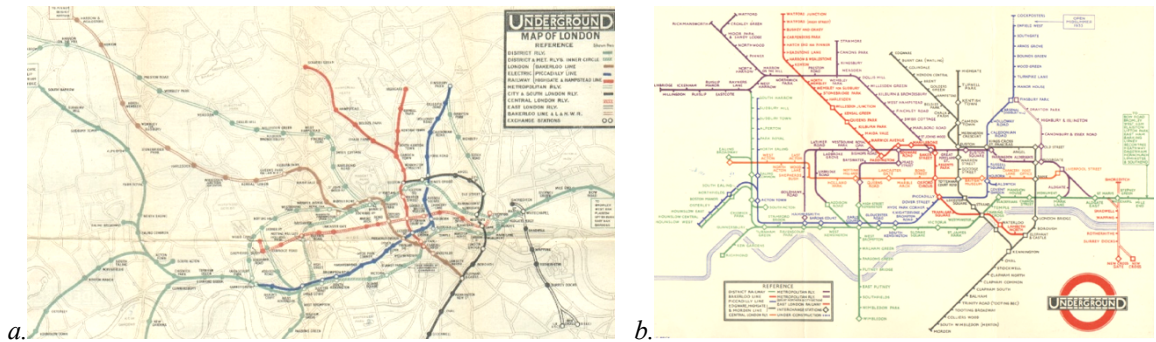

*a.*  *b.*

*Figure 2 London Underground map (a) before and (b) after Harry Beck redesign (1933)*

A similar advantage explains the appeal of force-directed layouts: they allow reading networks as geographical maps, despite the fact that network space is a consequence and not a condition of elements' positioning. In a force-spatialized visualisation there are no axes and no absolute coordinates, and yet the relative positioning of nodes is structurally significant. One can compare distances, gauge centres and margins, estimate density and often bring home interesting observations.

These insights, however, are not always easy to obtain. To observe relational structures the image of a network, one has to know not only where to look, but also how to make such structures visible. This is why the next section discusses at the same time how to read networks *and* how to improve their legibility.

## 2. How to read networks and make them legible

### The "jazz network" testbed graph

To exemplify our method, we wanted to use a "standard graph", but most testbed networks were too small for our purposes – for instance, the famous "Karate Club" of Zachary, 1977 contains only 34 nodes. It is easy to observe relational structures in networks of a few dozens or hundreds of nodes, but we wanted to show that VNA can also be applied to networks with several thousands of nodes. Inspiration came from another graph often discussed in the literature: the network of collaborations between jazz musicians produced by Gleiser & Danon (2003). As observed by McAndrew et al. (2014), "as a music form, jazz is inherently social" – because of the way jazz musicians are used to collaborate and form temporary ensembles – and thus particularly propitious to network analysis. Yet, Gleiser & Danon network contains only 1,473 nodes and is limited to the jazz bands that performed between 1912 and 1940 (making it difficult to interpret for the contemporary reader). We thus decided to produce an updated and expanded "jazz network" by drawing on Wikipedia's ontology. Here is the protocol that allowed us to obtain a graph of 5,985 nodes and 79,308 edges:

- We used Wikidata.org to extract
  1. All the 6,796 instances of "human" and the 976 instances of "band" with "genre = jazz". We thus obtained a list of individuals and bands that have a page in the English Wikipedia and that are related to jazz (mostly jazz musicians, but also jazz historians and producers). For each of them, we also collected (when available):
     o "birth year" (for individual) and "inception" date (for bands)

- "citizenship" (for individuals) and "country of origin" (for bands) – when multiple nations were available, we kept only the first one.
        - "ethnic group" and "genre" for individuals.
    2. All the 53 "subgenres" of the genre "jazz" and all the 396 "record labels" associated with the individuals and bands of the list above.
- We used the Hyphe web crawler ([hyphe.medialab.sciences-po.fr](hyphe.medialab.sciences-po.fr); Jacomy et al., 2016; Ooghe-Tabanou et al., 2018) to visit all the pages of the elements above in English Wikipedia and extract the hyperlinks connecting them.
- From the resulting network
    - We removed all the edges that did not have an individual or a band as one of their vertices (for reasons that we will discuss later).
    - We kept only the largest connected component (the largest group of connected nodes and edges), obtaining a network of 6,381 nodes (5,396 individuals, 589 jazz band, 346 record labels and 50 subgenres) and 85,826 edges.

## Positioning nodes

In section 1, we argued that the most important visual variable of VNA is the position of the nodes. Nodes that are more directly or indirectly associated, we wrote, *tend to* find themselves closer in the spatialised network. The caution introduced by "tend to" is crucial, because (as we will show in section 4), there is no strict correlation between the geometric distance in the spatialised graph and the mathematical distance (however defined) in the graph matrix. In VNA, we should not consider the exact position of any specific node, nor the distance between node couples, but the general grouping of nodes and the disposition of such groups. It is not the node positions that count, but the *density of nodes* and its variation. In particular, empty spaces should catch the eye of the observer.

In a continuum that goes from a set of disconnected nodes (a "stable") to a fully connected clique, the structure of a network is defined by the lumps and the hollows created by the uneven distribution of its relations. Since force-directed layouts would represent both extremes as circles filled with nodes placed at the same distance, everything that departs from this disposition is an indicator of structure. When analysing a spatialised network, therefore, we should look for shapes that are not circular – which indicate polarisation – and differences in the density of nodes – which indicate clustering.

Don't be too quickly discouraged, however, if your network looks like an amorphous tangle (a "hairball" as in network jargon). The legibility of network visualisations depends crucially on the choice of the spatialisation algorithm and its settings. Though all force-directed algorithms are based on a similar system of forces, their results may widely differ because of the specific way in which they handle computational challenges (in particular optimisations necessary to reduce calculations) and visual problems (in particular the balance between the compactness and legibility). What can, at first, be mistaken for a homogenous distribution of connections can sometimes derive from an unfortunate choice of spatialisation algorithm or settings.

This is why, among the many tools available for network analysis, we have a preference for Gephi ([gephi.org](gephi.org), Bastian et al., 2009). Developed expressly for network drawing, it does *not* treat spatialisation as an automated operation but offers a subtle control of visual variables. Among the force-directed algorithms our favourite is ForceAtlas2, because its performances on large networks and legibility of the results (cf. Jacomy et al., 2014). Alternatively, the JavaScript libraries Sigma.js (sigmajs.org) and Graphology (graphology.github.io) implement the same features in web technologies.

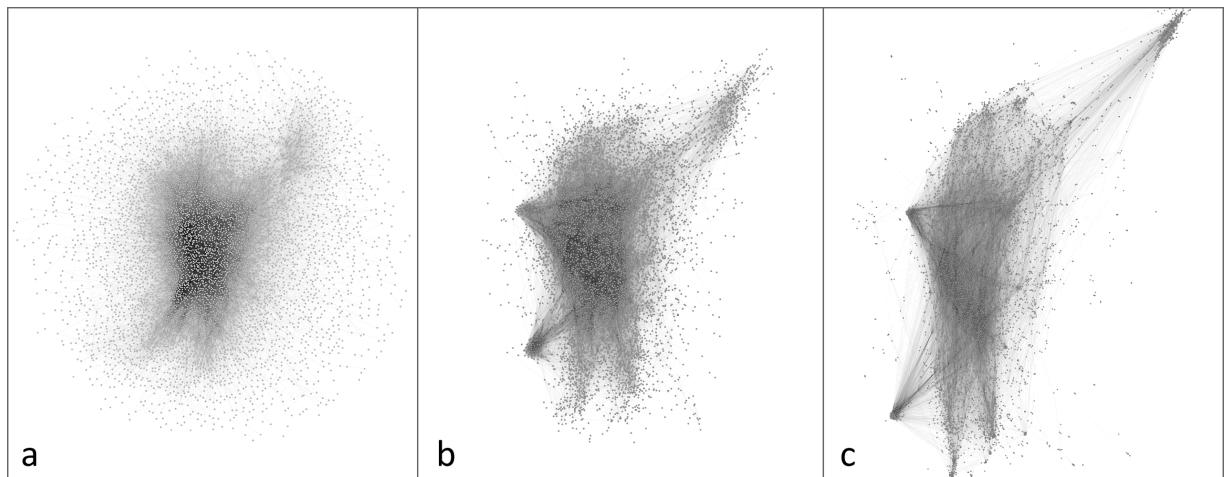

*Figure 3 The "jazz network" spatialised (a) with the algorithm proposed by Fruchterman & Reingold, 1991, (b) with ForceAtlas2 (with default parameters) and (c) with ForceAtlas2 with tweaked parameters for "LinLog mode" and "gravity"*

As an example, the image above shows how our network of jazz individuals and bands (for the moment, we are filtering out subgenres and record labels) looks like a hairball when spatialised with Früchterman and Reingold algorithm (considered as a milestone implementation of force-directed layout, see Früchterman & Reingold, 1991), but acquires a clearer structure when visualised with ForceAtlas2, particularly when two crucial parameters are adjusted.

The "LinLog mode" parameters tweaks the way in which distance is taken into consideration in the computation of attraction and repulsion forces. In default ForceAtlas2, both forces are linearly proportional to the distance (with inversed proportionality for attraction). However, as we will discuss in the last section of this paper, making the repulsion force proportional to the logarithm of the distance renders the clusters more visible – hence the presence of a "LinLog mode" in ForceAtlas2. "Gravity" on the other hand, is a generic force that pulls all nodes toward the centre. While it avoids disconnected nodes to drift infinitely far from the rest of the network, such a gravitational force interferes with the attraction-repulsion balance of force-directed layouts (a too high gravity packs all the nodes in the centre of the space). Activating the LinLog mode and setting the gravity to zero tends to make the clusters more visible, but also to produce a more scattered network (and require a longer time of computation to converge to an equilibrium, in particular for low-density networks). It is thus impossible to suggest a "catch-all setting" for these parameters. Recursively adjusting the spatialisation settings to the analysed network is crucial to make the relational structures visible (just as choosing the right chart and tweaking its visual properties is essential to make sense of a large data table).

### Sizing nodes and labels

Now that we have positioned the nodes of our network, in order to reveal effects of polarisation and clustering, we still have to make sense of what we see. To do so, VNA draws on two ancillary visual variables (Bertin, 1967): size and colour. Let's consider "size" first.

Tools like Gephi allow changing the diameter of the points representing the nodes according to a variable selected by the user. The degree (number of edges connected to a node) or, in directed networks, the in-degree (number of *incoming* edges) are classic choices, as they represent the most literal translation of visibility in networks. Being entirely relational, the degree can be computed for any network (and any directed network in the case of in-degree). Yet, when available, other non-relational variables could be equally interesting. For instance, we can change the size of nodes of our networks according to the number of visits that each of the related Wikipedia page received in 2017.

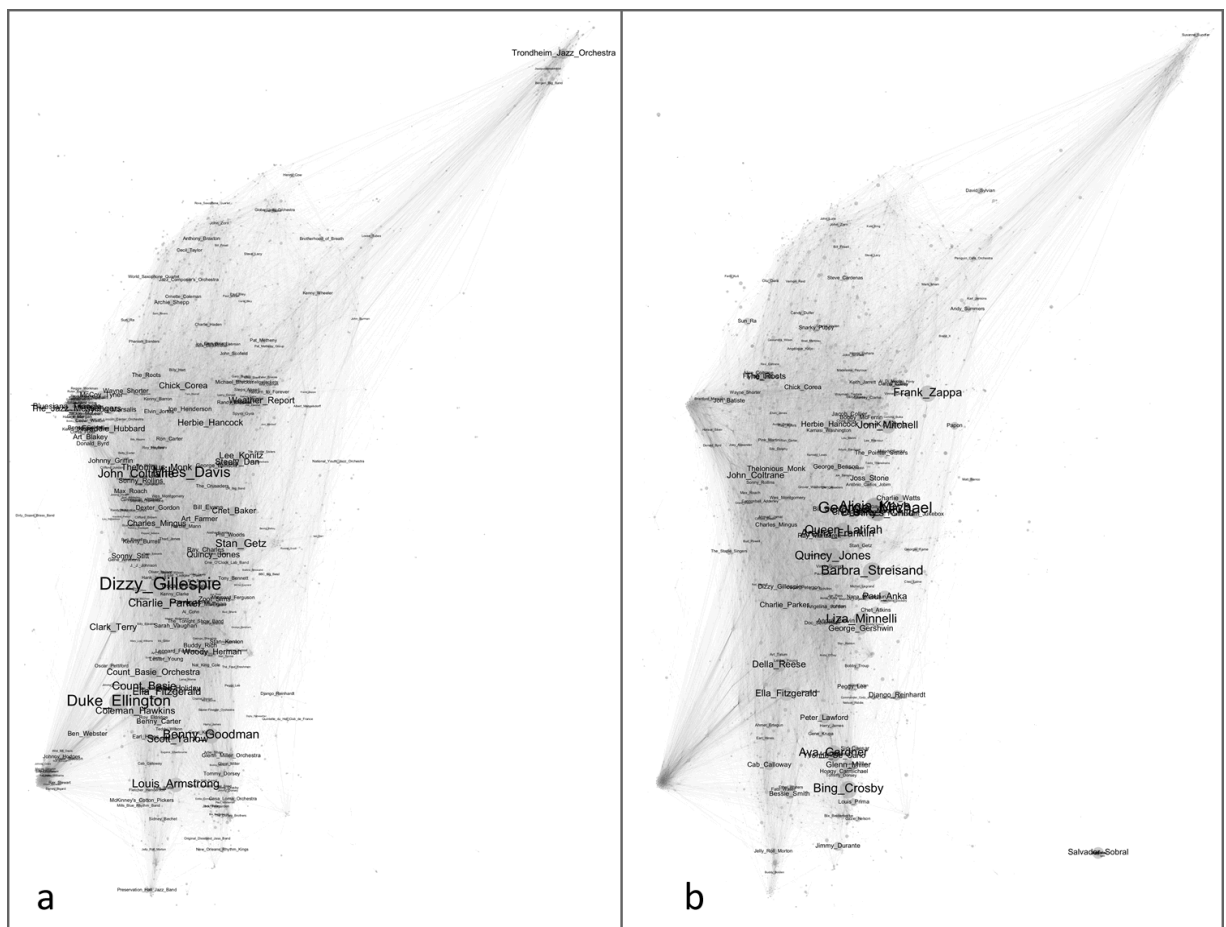

*Figure* 4 *The "jazz network" with nodes and labels sized according to (a) in-degree of the nodes; (b) number of page views of the related pages in the English Wikipedia.*

Note that in the figure 4, we have varied not only the size of the nodes, but also of their label (and even deleted all the labels smaller than a given threshold). This foregrounding operated through size is crucial in VNA because when working with networks with hundreds or thousands of nodes, inspecting all of them is clearly not an option. Changing label size (and dropping some labels), however, entails losing some information, and this is why using more than one scaling variable is always advisable.

Observing the labels of the most visible nodes, we can start to make sense of the factors that shape our network. Comparing the two images in figure 3, for example, it is possible to remark that nodes with a high in-degree tend to be on the left, while nodes with many page views (reminder: nodes represent Wikipedia pages) are rather on the right. Also, nodes with a high in-degree are all famous jazzmen (the top five being Dizzy Gillespie, Duke Ellington, Miles Davis, Benny Goodman and John Coltrane), while nodes with many page views seems to be pop-culture celebrities (the top five being George Michael, Alicia Keys, Barbara Streisand, Liza Minelli, Bing Crosby). In figure 4, changing the size of the nodes reveals a left-right polarization corresponding to a difference between a purer jazz lineage and the mixing with other genres.

This left-right polarisation, however, is not the most important. Indeed, if compared to a hairball circle, the network appears to stretch vertically much more than horizontally. How to interpret such a vertical polarisation?

### Colouring nodes

To investigate the vertical polarisation of our jazz network, we will add to position and size a third visual variable – colour. According to Jacques Bertin (1967), colour can be decomposed in at least two different variables: brightness (or value) which is better suited to represent continuous numerical variables, and hue, which is better suited to represent categorical variables (here we omit saturation, a third component of colour). VNA makes use of both.

Noticing at the bottom names such as Louis Armstrong, Duke Ellington and Bing Crosby and at the top Chick Corea, Weather Report and Frank Zappa, we can hypothesise that the vertical polarisation of our network is connected to time and in particular to the period in which the different actors were most active in the jazz scene. While such information is not available in our network, we do have the year of birth and of inception of individuals and bands and we can project them on the network using a scale of brightness going from black (for the oldest actors) to white (for the newest) – see figure 5a.

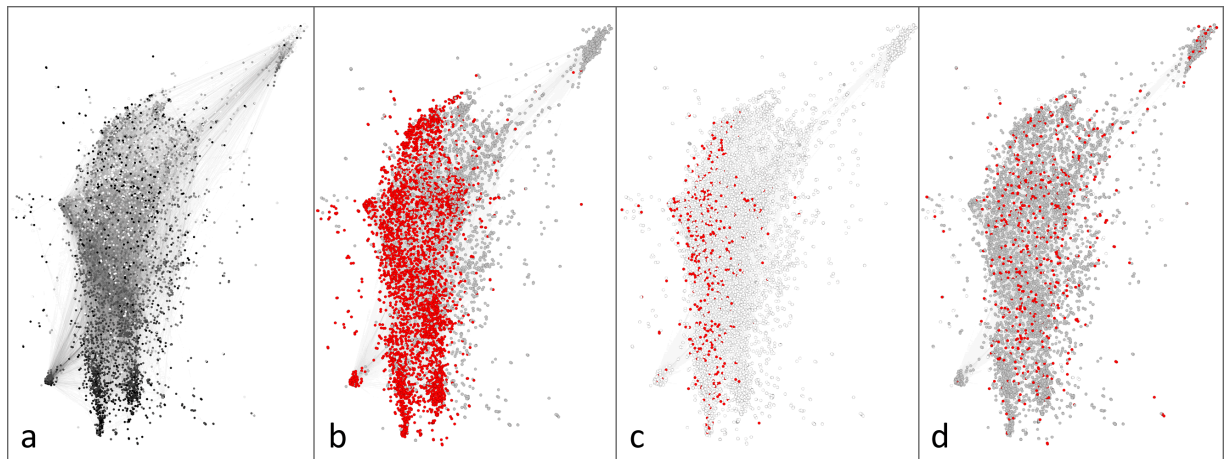

*Figure* 5 *The "jazz network" with nodes coloured according to*
*(a) the year of their birth or inception (from dark for older individuals and bands to white for newer);*
*(b) their nationality (red for US, grey for all other countries, white for not available);*
*(c) their ethnic groups (red for African American, grey for other ethnic groups, white for not available);*
*(d) their genre (red for women, grey for men, white for not available or others)*

Figure 5a seems to confirm our hypothesis that the vertical polarisation corresponds to time. While the separation is not complete, darker nodes are more present at the bottom of the image and brighter nodes at the top.

In figures 5b, 5c and 5d, we relied on hue (red, grey and white) to observe how different categories distributes in the network. Figure 4b and 4c are dedicated respectively to nationality and ethnic group. While they are difficult to interpret alone, together they suggest an interpretation. Figure 4b, reveals unsurprisingly that jazz is primarily an American genre of music (remember, however, that we relied on English Wikipedia to build the network), but it also shows that most non-American actors (in grey) tend to be on the right of the image.

Similarly, figure 4c shows that while most nodes are not qualified, the only ethnic group that stands out is African American (again not surprisingly knowing the history of the genre). The nodes representing African American actors (in black) are everywhere in the network, but slightly more to its left than to its right. Both observations seem to confirm the interpretation we got from figure 4, that the horizontal polarisation is loosely connected to the "purity" of the jazz genre.

To be sure, not all variables will turn out to be connected to the visual structures of the network. In figure 5d for example, we show how genres are completely mixed in our network, in a way that suggests that at least in this field genre does not produce a relational fracture (but notice how men are significantly more numerous than women).

Using force-directed spatialisation to determine the position of nodes and size and colour to project various variable on our visualisation, we have identified two different polarisations of our network: time, determinant to the general structure and varying bottom-up, and "genre purity", less determinant to the structure and varying from left to right. This configuration is distinctive of this network and is not to be expected in every network. Other combination of two or more polarisations can give shape to different visual structures (see Boullier, Crépel & Jacomy, 2016).

One might be tempted to interpret the spatialisation in terms of axes: the vertical axis would be time and the horizontal axis would be genre purity. This would be a mistake. In spite of superficial similarities, force-vector algorithms work differently from dimensionality reduction techniques like correspondence analysis (Ter Braak, 1986). Force-vector layouts do not produce coherent axes. They consider space as isotopic (the same in every direction) and implement no concept of "top/bottom" or "North/South/East/West". As a consequence, polarization is generally not coherent across different clusters: the same variable might spread left-to-right in one cluster and top-down in another.

## Naming poles and clusters

So far, we have looked only at the *poles* of our graph, not at its clusters. We have considered the shape of the network, but not the different zones of density produced by the disposition of nodes. In VNA clusters are defined as regions that gather by many nodes closely packed together and surrounded by areas with a much sparser density (the "structural holes" of Burt, 1995).

In the jazz network, the only easily identifiable cluster is the one located at the very top right of the image and whose most visible node is the Trondheim Jazz Orchestra (see figure 3), which contains a group of mostly Norwegian musicians, most of which are members of the Orchestra. The other clusters of our network are more difficult to identify and interpret. To do so, we present in this paper two advanced techniques for visual network analysis. These techniques facilitate, but do not replace the basic operation of thoroughly examining the density and reading nodes labels and qualification (when available) to make sense of why some groups of nodes are more closely connected than others.

The first technique is not available in Gephi but can be performed by a script run in an online tool called Graph Recipes (tools.medialab.sciences-po.fr/graph-recipes). Using such a script (available at https://github.com/tommv/ForceDirectedLayouts as all the scripts that we used in this article), we transformed our network in a heat map in order to make the differences of density more salient (see figure 5).

The second technique entails qualifying the different areas of the network using "qualifying nodes". This technique consists of adding to the network a new set of nodes that do not influence the spatialisation but can be used to make sense of it. In our example, we used the subgenres of jazz (according to Wikidata) and the record labels associated with the artists and ensembles of our network. To make sure that these qualifying nodes do not influence the layout, we used a "double spatialisation". We first spatialised a network containing only the nodes representing the individuals and the bands. We then froze the position of these "primary nodes", added the subgenres and record labels and run the spatialisation algorithm a second time on the qualifying nodes only. A last detail: though the Wikipedia pages related to the subgenres and record labels have hyperlinks connecting them, we have removed these edges from our network, so that the qualifying nodes are only position according to their connections to the primary nodes (and not according to the connections between themselves).

After the double spatialisation, the qualifying nodes can be used to suggest labels for the clusters of the networks in which they end up being located. To complete our visualisation, we worked with a jazz expert (Emiliano Neri, whom we heartfully thank for his help), to drop most primary and qualifying labels and keep only the most significant according to his expertise.

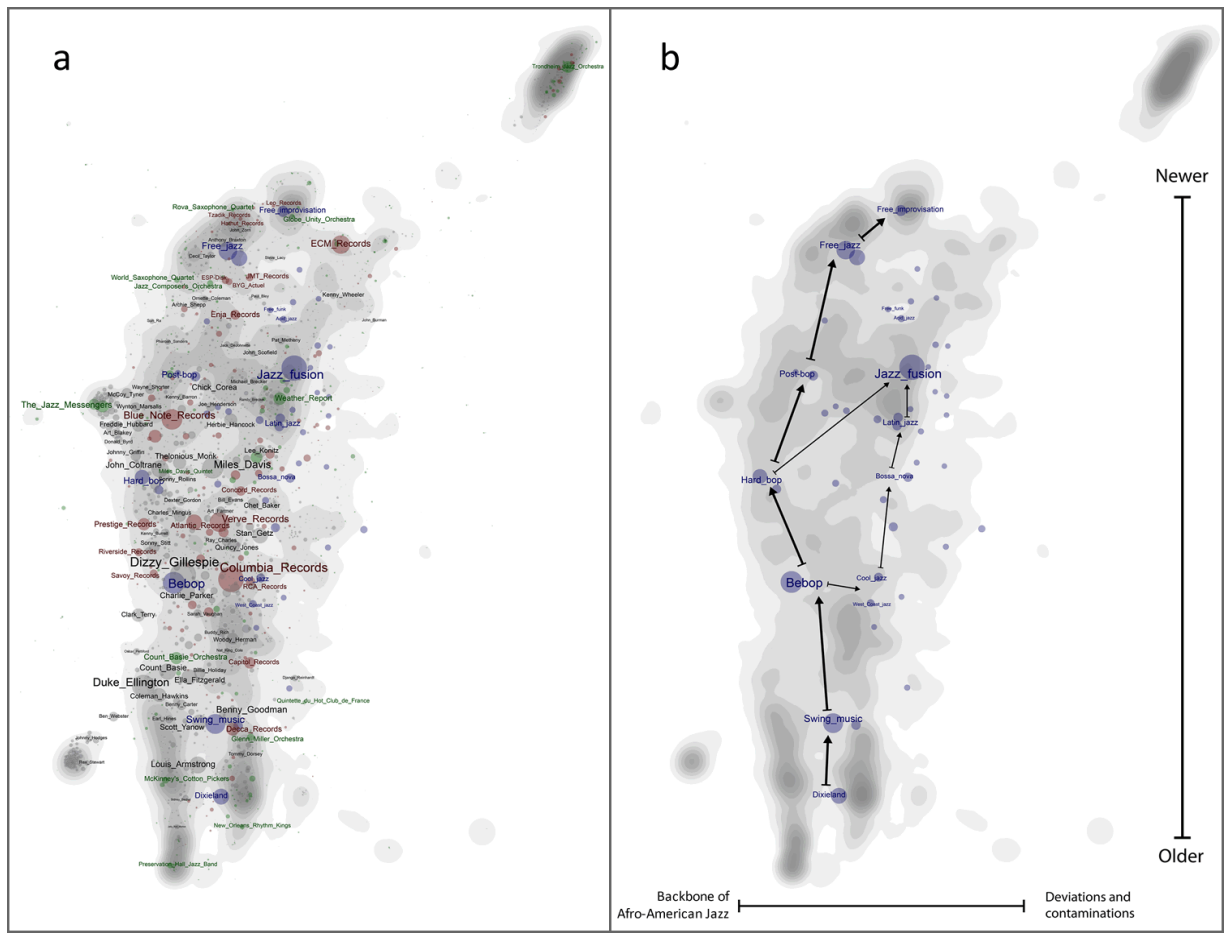

*Figure 6 The "jazz network" with (a) the labels of the most salient node of each type (grey for individual, green for bands, blue for subgenres and red for record labels) and (b) the identification on the structure of the network in terms of the evolution of the jazz musical language.*

### Interpreting the position of nodes and clusters

Now that we have decided how to spatialize the network, and hierarchized the visualisation to keep only the most important landmarks, we can try to make sense of both its overall structure and the position of its most important nodes. As we will argue in the next section, it is a distinctive advantage of VNA that it allows observing global patterns and local configurations in the same visual space.

In figures 6 and 7, one can observe (moving from the bottom to the top of the image) the development of the jazz musical language. This evolution occupies the left of the image and starts from *dixieland* and *swing music* and progresses to *bebop*, *hard bop*, *post bop* and finally to *free jazz* and *free improvisation*. From this backbone of Afro-American jazz, depart on the right of the charts deviations (such as the *cool jazz* and *west coast jazz*) and contaminations with other genres (such as *bossa nova*, *latin jazz* and later *jazz fusion*).

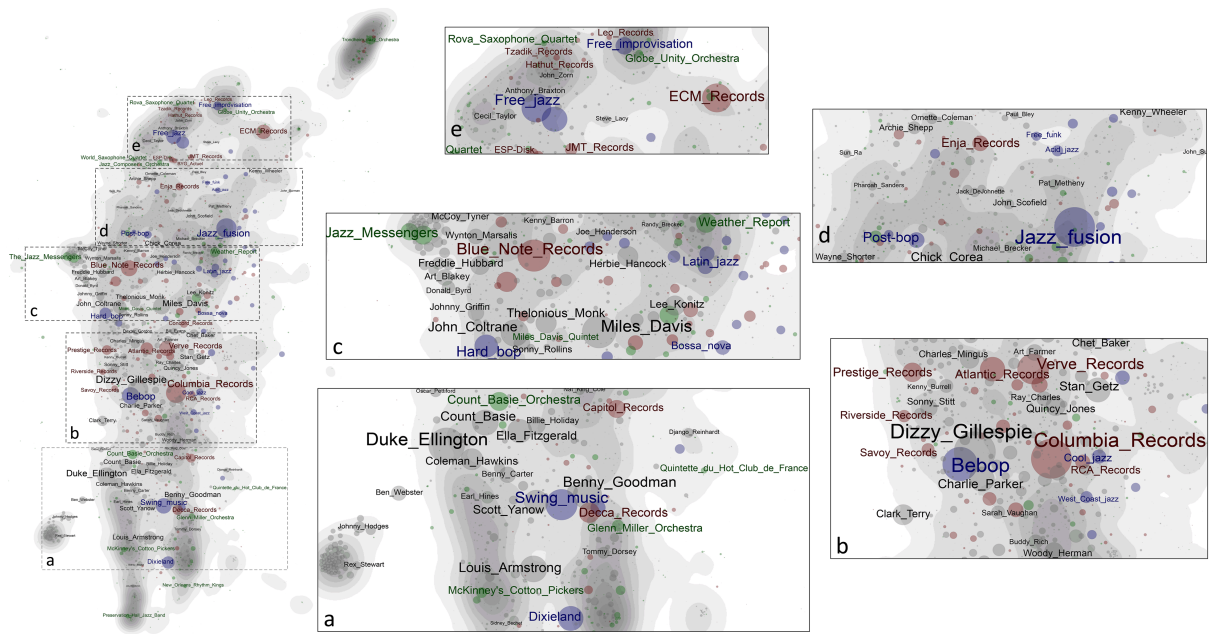

*Figure 7 Mosaic providing a zoom on the different regions of the "jazz network"*

[7.a] The bottom of the image corresponds thus to the early years of the genre and is marked by Decca Records, a label which dominated the jazz scene in the 1930s and 1940s, and Capitol Records, also particularly active in the 1940s. The region of *dixieland* and *swing music* is split in two parallel clusters (already identified by Glaiser et al., 2003): to the right, "white big bands" gathered around *Tommy Dorsey*, *Glenn Miller* and *Benny Goodman*; and to the left the "black big bands" gathered around *Louis Armstrong*, *Coleman_Hawkins*, *Count Basie* and, *Duke Ellington*. This last bandleader is also at the origin of the smaller cluster to the bottom left, constituted by the members of its orchestra. Famous vocalists such as *Ella Fitzgerald* and *Billy Holiday* are positioned toward the centre because of the large number of their collaborations. More to the right, is *Django Reinhardt*, the Romani guitarist, whose isolated position is justified by his living in in Europe.

[7.b] Shifting up toward the *bebop*, many new record labels emerge such as *Prestige*, *Riverside*, *Savoy*, *Atlantic*, and more importantly *Verve* and *Columbia* which were destined to impose themselves in the jazz scenes for years to come. Very close to the node representing *bebop*, one can find (not surprisingly) the trumpeter *Dizzy Gillespie* and the saxophonist *Charlie Parker*, who were among the most influential artists of this new style, and the vocalist *Sarah Vaughan* who collaborated with both. In a more bridging position are *Woody Herman* and *Clark Terry*, whose long careers spanned between *swing* and *bebop*.

[7.c] Moving upward, the increase in the number and dispersion of nodes illustrates the growing diversification in jazz language in the 1950s. On the one hand (on the left of chart), *bebop* evolves into *hard bop*, thanks to the *Blue Note* record label and to musicians such as *Charles Mingus*, *Sonny Rollins*, *Thelonious Monk* and *Art Blakey*. This last bandleader is at the origin of the important ensemble of the *Jazz Messengers*, which creates a little cape on the left of the map and which acted as an incubator for talent, including *Freddie Hubbard*, *McCoy Tyner* and *Wynton Marsalis*. On the other hand (on the right of the chart), the experiences of west coast jazz and cool jazz evolve through the contamination with styles from Latin America, giving birth to *bossa nova* and *latin jazz*, popularized in the US by influential figures such as *Stan Getz* and *Quincy Jones*. *John Coltrane* and *Miles Davis* occupy the centre of this region (and of the whole graph) for the crucial role they played in bridging all these experiences.

[7.d] In the 1960s, the contaminations observed in the centre-right of the chart turn toward rock and funk music as well as their use of electric instruments and amplifiers, originating the so-called *jazz fusion*. Musicians such as *Chick Corea*, *Herbie Hancock*, *John Scofield* and *Pat Metheny*, as well as the group *Weather Report*, play a crucial role in this experience. At about the same time, and with

connections assured by artists such as *Joe Henderson* and *Michael Brecker*, *hard bop* develops into *post-bop* thanks to musicians such as *Wayne Shorter* and *Elvin Jones*.

[7.e] In the 1970s, experiences of radical improvisation developed in the previous decades conquered the musical avant-garde, giving birth to *free jazz* and *free improvisation*. Initiated by musicians such as *Sun Ra*, *Cecil Taylor*, *Archie Shepp* and *Ornette Coleman*, this style has been developed by *Anthony Braxton*, *John Zorn*, *Evan Parker* and many others. Interestingly, this genre seems to be edited particularly by European record labels such as *JMT* and *ECM*. This last record label is also the bridge that connects the relatively marginal cluster of the Scandinavian jazz (at the top-right of the figure) to the rest of the maps.

## 3. What type of analysis is visual network analysis?

In the introduction we mentioned that visual network analysis is particularly suited to the medium-sized networks (ranging from hundreds to thousands of nodes) often found in social and biological phenomena. In this section we will elaborate on this idea, demarcating VNA both from older styles of network visualisation and from analytical approaches that avoid visualisation all together.

### Not too small – diagrammatic vs topological visualisations

Besides introducing the basic techniques of VNA, the example of the jazz network should have provided an illustration of the specific sense in which this analysis is a method for *reading* networks. This is important because, while force-directed algorithms have always been employed to improve networks legibility, the meaning of this "legibility" has evolved over time.

Historically, force-directed layouts were introduced to satisfy aesthetic criteria like "minimizing edge crossings" or "reflecting inherent symmetry" (Früchterman & Reingold, 1991; Purchase *et al.*, 1996; Purchase, 2002). In the word of Jacob Moreno "the fewer the number of lines crossing, the better the sociogram" (1953, p. 141). We call this approach *diagrammatic* as its objective is to prevent visual cluttering (e.g. by reducing node and edge occlusion) and increase reading comfort (e.g. by optimising the occupation of space). In this perspective, network visualisations are similar to flow charts and reading them means following their paths and observe which nodes are connected.

This diagrammatic reading, however, is only possible for small networks for the intricacy of visualisations grows so quickly that even graphs with a few hundreds of nodes are impossible to read in this way. This does not mean, however, that force-directed layouts lose all their interest. Quite the opposite, as we have illustrated with the jazz network. Though introduced to serve diagrammatic objectives, force-directed spatialisation revealed an interesting by-product: their capacity to make network structures visible. When the balance of forces is reached, clusters *tend to* appear as denser gatherings of nodes and edges; structural holes *tend to* look like sparser zones; central nodes move *towards* middle positions; and bridges are positioned *somewhat* between different regions.

We will call this second perspective *topological* as its objective is to turn topological structures into visual characteristics. Diagrammatic and topological perspectives coexist in practice and are often confused in literature as some aesthetic features (symmetry for instance) may be desirable both *per se* and as a translation of topology. Nevertheless, the two approaches come from different traditions and have different aims – *graph drawing* is rooted in algorithmics and focuses on single paths through the network, while *network visualisation* stems from information design and is interested in relational structures. For instance, replacing edges with a density heat map (as we did in fig. 6) makes sense in a topological perspective because it helps revealing clusters but is absurd from a diagrammatic viewpoint for it defeats the objective of following the paths.

Ultimately, the two perspectives serve different needs. A diagrammatic stance suits small networks, whose configuration is simple enough to be qualitatively appreciated, while a topological attitude is more appropriate for larger networks, where pattern detection is preferred. This explains why, in the last few years, the attention of scholars has gradually shifted from the diagrammatic to topological approach.

Diagrams, favoured in the early years of network visualisation – at a time when large networks were expensive to collect and compute – are becoming obsolete when confronted to the growing size of relational datasets (Henry *et al.*, 2007). Reviewing an assessment of spatialisation algorithms by Purchase *et al.* (1996), Gibson *et al.*, (2013) note for instance:

> What is interesting though is the type of tasks she [Helen Purchase] asked her users to complete. These were finding shortest paths, identifying nodes to remove in order to disconnect the graph and identifying edges to remove in order to disconnect the graph … It is unclear as to if this type of accurate, precise measurements are a typical analysis tasks for graphs with hundreds or thousands of nodes …. If those kinds of tasks become infeasible due to the volume of nodes and edges then the better layouts should support the user for a different set of tasks … Another aim for layout should then be to support users in tasks concerned with overview, structure, exploration, patterns and outliers (pp. 27, 28)

### Not too big – visual ambiguity for exploratory data analysis

In presenting our analysis of the jazz network, we have described the sequence of transformations imposed on our graph to make it readable. For the sake of simplicity, we presented such sequence as linear and orderly, as if we knew from the beginning how to stack the different operations and how to set the different parameters. Of course, this was not the case and our actual enquiry entailed many trials and errors, as well a lot of backs and forth between different visual variables and their parametrization. At some point we even realised that our original data collection was flawed and had to be renewed (because we initially treated jazz ensembles as "qualifying nodes", while in the analysis their role turned out to be akin to that of jazz musicians). This type or iteration is very common in visual networks analysis, which cannot be carried out without a continuous switch between data and visualisation, zooming and panning, selecting and filtering.

VNA is in this sense a form of "exploratory data analysis" (Behrens & Chong-Ho, 2003). Introduced by John Wilder Tukey (1977), EDA covers all the operations that researchers carry out to make sense of their data, before (and in order to) formulating clear hypotheses or findings (the so-called "confirmatory data analysis"). VNA facilitates exploration because in contrast with graph metrics it conserves each node and edge as a separate element. This allows to interpret networks patterns by observing single nodes (e.g. when we relied on "subgenre nodes" to identify the clusters of our jazz networks) and conversely to appreciate the role of single nodes by considering their relative position (e.g., when we observed the crossroad position of John Coltrane and Miles Davis).

The effort to maintain the singularity of nodes and edges, however, precludes the use of VNA on graphs of millions of nodes and edges. When dealing with massive datasets harsher forms of aggregation and dimensionality reduction may be necessary. And yet, as we tried to prove with the jazz network, visual networks analysis scales up nicely to graphs of thousands of nodes, making it capable to cover many medium size social and natural phenomena.

Still, for many network analysts, it is not the scalability of network visualisations that is problematic, but their lack of exactitude. To them, it makes no sense to use an ambiguous method of analysis (drawing) when a more precise one exists (calculus). Why, for instance, spending time to figure out which nodes occupy the unclear centre of a fuzzy community, when most graph statistics offers centrality indicators by the dozen?

Though this may sound odd to the mathematical ear, ambiguity may in fact be a *strength* of network visualisations. As for all statistical indicators (Desrosières, 1993), the precision of graph metrics derives from a reduction of the information available. It is the very nature of numerical indicators to discard much of the complexity of the empirical phenomena and focus on the few dimensions that can be precisely quantified.

This *reduction to exactitude*, which constitutes the greatest asset of statistical metrics, can be a drawback in the early stage of scientific investigation, when the definition of the research questions is still underway and the mastery of research corpora still tentative. As long as the separation between

"information" and "noise" (or "measure" and "error", if you prefer) remains unclear, premature efforts to *clean the picture* risk to *cut* observation along precise but fallacious lines. In early stages, researchers should respect the inherent ambiguity of their subjects rather than imposing them a premature and artificial ordering. Hence the interest of exploratory data analysis and of maintaining margins of approximation, at least in the early appraisal of data. In Tukey's own words:

> "Far better an approximate answer to the *right* question, which is often vague, than an *exact* answer to the wrong question, which can always be made precise." Data analysis must progress by approximate answers, at best, since its knowledge of what the problem really is will at best be approximate. It would be a mistake not to face up to this fact, for by denying it, we would deny ourselves the use of a great body of approximate knowledge (Tukey, 1962 p. 14, original quotes and emphasis).

Maintaining margins of ambiguity is particularly important in human and social sciences. Because of the complexity of their subjects, many researches in these fields cannot bear the degree of exactitude implied by confirmatory statistics. If many human and social scientists avoid mathematical tools, it is not because they do not understand them, but because their abstraction is at odds with the singularity, reflexivity and richness of human phenomena. Johanna Drucker (2011), for example, argues that standard statistical charts convey a purity that is unrealistic for most social categories.

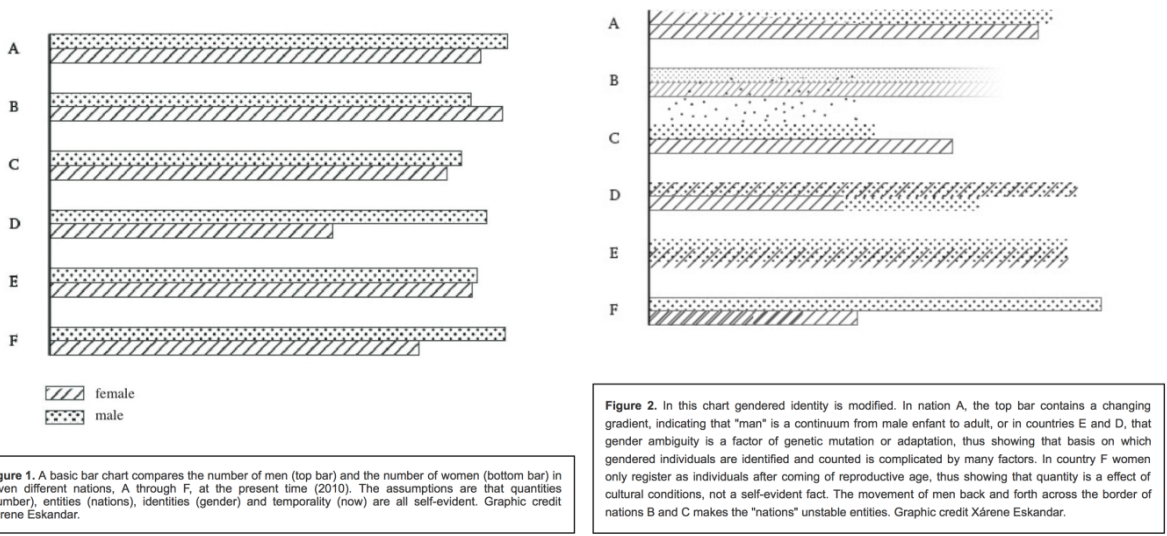

*Figure* 8 *A classic statistical chart (left) and its redesign with some of the original ambiguity (right) (original images and captions from Drucker, 2011)*

This is one of the reasons why network charts are increasingly popular as ways to explore complex subjects: their visual ambiguity mirrors some of the empirical ambiguity of the phenomena they represent. Where are the frontiers of a group? What is the overlap between two clusters? Which nodes occupy the centre of a community? What are the bridges between different relational regions? None of these questions has an exact answer in social or ecological research and accordingly, network visualisation imposes no answer, often in contrast to network metrics. Where, for instance, community-detection algorithms tend to generate clear-cut and non-overlapping partitions, force-directed spatialisation reveals zones of different relational density but with blurred and uncertain borders.

In force-directed layouts, the structural holes between clusters appear as gradients, like a valley between two hills. Like cartographic representations, networks visualisations offer orientation without imposing clear-cut boundaries. In the same way in which it is possible to locate and chart hills and valleys even if it is impossible to define the exact point where one ends and the other begins, so VNA conserves the inherent vagueness of concepts such as clusters, centres, fringes and bridges. Network metrics (and network models) are great tools to test for relational hypotheses, but network maps can be just as useful when the problem is to explore uncertain phenomena. Not *despite* their ambiguity, but *because* of it.

Because they are problematic, graph visualisations encourage researchers to problematize their observations and encourage an enquiring attitude (Dewey, 1938).

## 4. Toward a measure of spatialisation quality

Effective in practice, the techniques presented above remain conceptually underdeveloped and their efficiency has not been properly evaluated. As observed by Bernhard Rieder and Theo Röhle: "tools such as Gephi have made network analysis accessible to broad audiences that happily produce network diagrams without having acquired robust understanding of the concepts and techniques the software mobilizes. This more often than not leads to a lack of awareness of the layers of mediation network analysis implies and thus to limited or essentialist readings of the produced outputs that miss its artificial, analytical character" (Rieder & Röhle, 2017). Force-directed algorithms, we argued are precious exploratory tools but are not exempt from bias.

While we have praised the value of approximation in exploratory data analysis, ambiguity is not arbitrariness and good network visualisations should separate the uncertainty inherent to the data from the distortions due to the particular graph embedding technique. The next figure illustrates the problem. A chain of four nodes can be drawn in a way that is directly justifiable by its structural properties – nodes B and C are twice as close A and C or B and D, because they are connected twice as directly. A clique of four nodes, however, cannot be properly designed– since all nodes are equally connected they should be at equal distance one from another, which is impossible (unless, of course, if the nodes are all positioned in the exact same point). The example illustrates the distortion necessary to fit the n-dimensions of a graph adjacency matrix in the two dimensions of a computer screen (or piece of paper). Force-driven layouts, like any other dimensionality reduction techniques, entail distortions and losses of information.

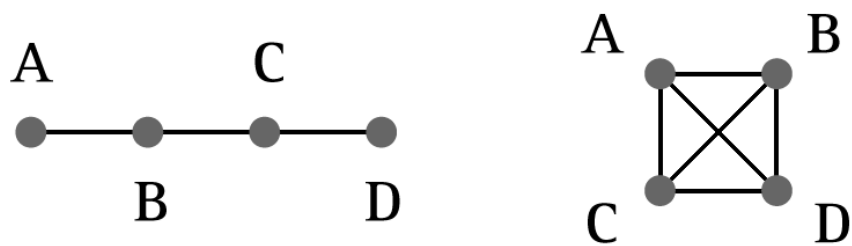

*Figure 9 An exact network spatialisation (left)   –   A necessarily skewed network spatialisation (right)*

Separating the distortions produced by the need to fit a multidimensional object in a two-dimensional space from the intrinsic elusiveness of relational phenomena, however, is far from easy, because of the difficulty to find a ground truth against which to evaluate the performance of force-directed layouts. The most natural standards to evaluate geometric distances in the visualisation are mathematical distances. Unfortunately, two major problems arise. Not only there are multiple distances that disagree with each other, but we have evidence that none of those corresponds to the specificity of force-directed layouts.

In figure 10, We compare the Euclidian distance between each couple of nodes of the jazz network as spatialised by ForceAtlas2 (LinLog and gravity 0) with two major topological distances: the length of shortest path (geodesic distance) and the mean commuting time. This last quantity is defined as the average number of steps that a random walker, starting from one node, takes to reach the other for the first time and then go back to the starting node (Fouss et al., 2007). The Euclidian distance is somewhat correlated with the geodesic one, as expected, but the variability is very strong, because the position of a node is influenced by its distances to many nodes. There is almost no correlation with the mean commuting time, as random walkers can drift considerably far away from even a neighbouring node, especially when nodes' degrees are high.

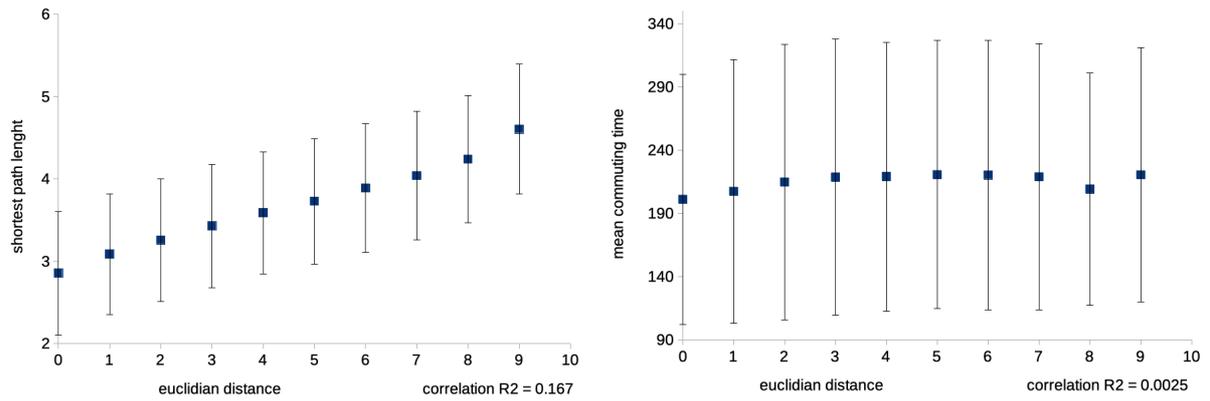

*Figure* 10 *scatter plots showing, for the jazz network that the Euclidean distance in the spatialised network (ForsceAtlas2, LinLog mode, gravity = 0) is poorly correlated to both the shortest path and the random walk distances (respective R2 : 0.167 and 0.0025). The plot represents the mean and standard deviation as error bars) of the respective distances for the binned Euclidean distances « i » ( i=int(distance / distance_max).*

Considering the distance between nodes has also the disadvantage of discounting the topological nature of network spatialisations, which aims at representing the underlying relational structures rather than at showing the connections between individual nodes. As of today, we cannot offer a fully convincing answer to the problem of measuring the quality of force-directed spatialisation. We can, however, formulate the question as clearly as possible, framing the problem in a way that helps finding a solution and, hopefully, spark the curiosity of mathematicians.

We will start by reformulating the question of spatialisation quality as: *how close are force-directed algorithms to producing an optimal disposition of nodes*? Interestingly, even though we have no precise definition of "optimal", we have two reasons to suspect that current algorithms may be near-optimal. The first is the pervasiveness of spatialisation techniques. Not only they have been used for more than a decade with no major modifications, but they have also expanded to other areas. Indeed, dimensionality reduction algorithms in multivariate variable distribution, like t-SNE (van der Maaten & Hinton, 2008) and UMAP (McInnes et al., 2018), are implicitly building networks and spatialising them. The way they minimize entropy by gradient descent (GD) bears a striking resemblance to force-directed layouts. Both are iterative "relaxation" techniques converging to an approximate equilibrium and both are meant to optimize a function, which is explicit for GD and implicit for network spatialisation (and roughly corresponding to the energy of the system). The increasing success of T-SNE and UMAP suggests that collective intelligence has not found better than these quite similar techniques to produce interpretable visual objects.

The second reason is Andreas Noack's work on the LinLog algorithm. In his thesis (Noack, 2007) he proposes a layout quality metric called "normalized atedge length", corresponding to the total geometric length of the edges divided by the total geometric distance between nodes and by the graph density. The smaller is the value of this metric, the better the layout has succeeded in producing compact and separated clusters, for its numerator increases when connected nodes are close, and its denominator decreased when disconnected nodes are far. Note that, in a sense, this is what all force vectors try to accomplish, because attraction only applies to connected nodes while repulsion applies to all nodes.

Regrettably, the metric does not set an optimum expectation level and does not quantify the amount of bias due to the constraints of reducing dimensionality. It can, however, be used to compare different spatialisations and, as shown by Noack, to prove that the best results are produced by force-directed layouts employing a linear force of attraction (i.e. linearly proportional to the distance of nodes) and a logarithmic force of repulsion. Such is the principle of Noack's "LinLog" algorithm (Noack, 2007), often considered as the gold standard of spatialisation quality.

In a later paper (Noack, 2009), Noack also demonstrated, for a very simple network, how his metric (the normalized atedge length) is mathematically equivalent to modularity, a clustering quality metric used for community detection (Newman, 2006). This result provides evidence that the LinLog algorithm may be close to optimum in the task of translating mathematical community-structure into a visual clustering. It also indicates that the problem of "normalized atedge length" minimization is probably NP-complete, as is the problem of maximizing modularity (Brandes et al., 2006). This indicates that it may be hard to outperform the iterative convergence of force-directed layouts by using a deterministic approach.

The correspondence between structural and visual clustering suggests a simple heuristic to gauge network layouts. Good layouts should suggest a visual partitioning of a network roughly corresponding to its mathematical clustering. As an illustration, the following figure reports the Jaccard similarity between the geometric cuts computed by k-means and its structural cuts computed by Louvain modularity (Blondel et al., 2008).

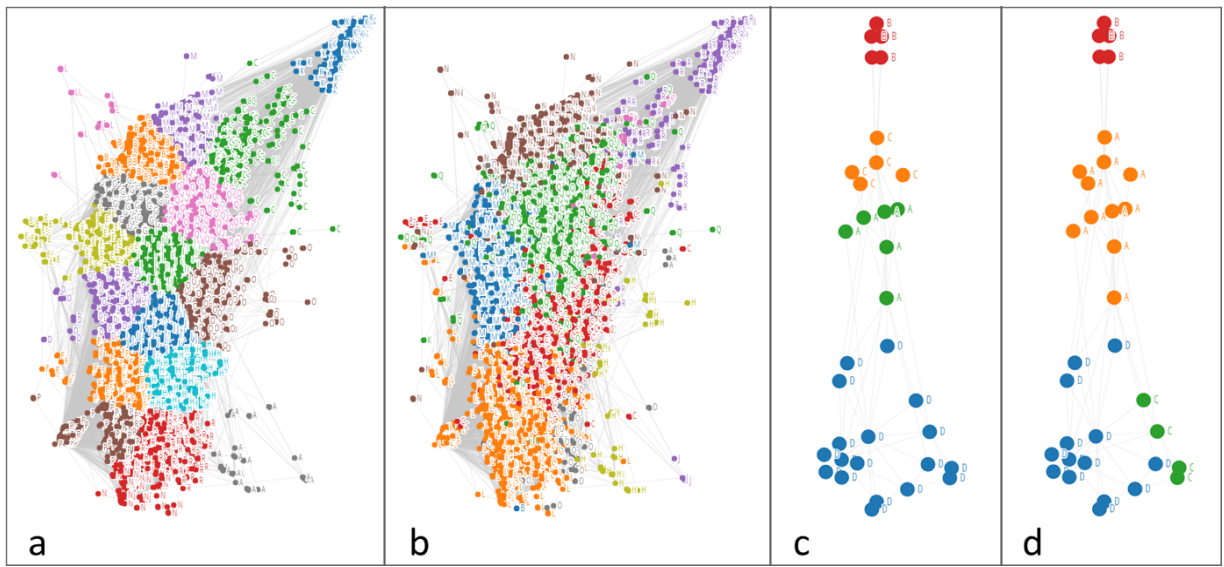

*Figure* 11 *Comparison of the clustering cuts identified by k-means (a and c) and Louvain modularity (b and d) on the jazz network and the karate club network.*

The bar chart below confirms the insights of the images above by proposing a more systematic comparison of the networks cuts identified by Louvain modularity and k-means (using the number of classes found by the Louvain algorithm in both cases) on three different networks (the jazz network, the karate club and a clique added for comparison) and four layouts (Force Atlas with gravity zero and linlog mode, default Force Atlas, default Früchterman & Reingold and a random layout added for comparison). Each bar in the diagram represents the correspondence between the cuts of the two clustering algorithms through a simple Jaccard similarity. For each pair of partitions, we compared how similar they were by computing the Jaccard index of the set of pairs of nodes that are in the same partition. This process can be unpacked this way:

1. For a given network and a given partition of the nodes in *k* different classes *C*
2. We build the set *S* of all pairs of nodes *($N_i$, $N_j$)* where the classes *$C(N_i)$* and *$C(N_j)$* are the same:
   *$C(N_i) = C(N_j)$*
   In other terms, this is the set of the node pairs that define the clusters.
3. To compare the two partitions *a* and *b* of the same network, we compare the sets *$S_a$* and *$S_b$* with a Jaccard index: how many pairs *($N_i$, $N_j$)* are in common, over how many pairs are in either or both of the sets.

The Jaccard index has a value of 0 if the partitions have no node in common, and a value of 1 if they are exactly the same. Comparing the pairs of nodes has the benefit of not requiring matching each cluster of partition *a* with a cluster of partition *b*, which cannot always be done in a meaningful way.

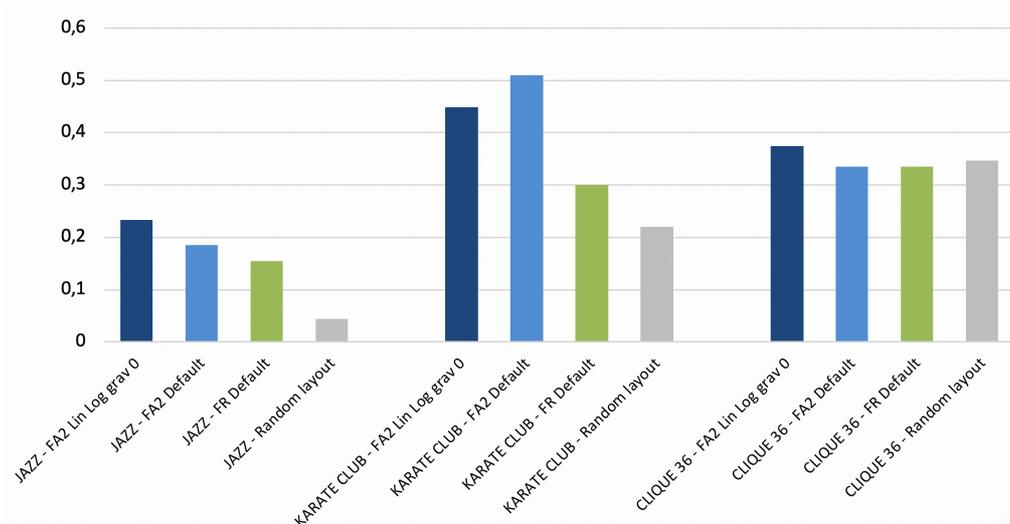

*Figure* 12 *Bar chart of the Jaccard similarity between the partitions identified by Louvain modularity and k-means on the jazz network, the karate club network and a clique of 36 nodes.*

This simple comparison between modularity and k-means partitions works well on highly clustered networks, but not on networks that are more structurally ambiguous and exhibit some polarization but not clear-cut clusters, which is why the karate club network exhibits a higher cut similarity than the jazz network. This not surprising. If, as we argued, the value of spatialisation lies in its capacity to conserve and render ambiguity, such value will be poorly captured by a measure based on cuts.

Searching for a spatialisation quality metric is a case of experimenter's regress (Collins, 1975), a situation where we face a dependency loop between theory and evidence. To avoid the confirmation bias, it is crucial to remember that the LinLog is an *empirical* gold standard, whose theoretical soundness still needs confirmation. We are not sure that Noack's "normalized atedge length" is the metric that should be minimized, and we have no proof that the LinLog is the best possible approach to minimize it. All we know is that the "normalized atedge length" is a reasonable definition of spatialisation quality and that, among the existing layouts, LinLog is the one that delivers the best results according to it (while also being the one that makes structural clustering most easily observable).

To provide a solid mathematical ground for visual network analysis, however, we still need a spatialisation quality metric independent from the current state of algorithms and based on a deeper theoretical ground. As we discussed above, we have reasons to think that such metric exists; that it is different from (but related to) existing metrics like the geodesic distance; and that existing force-oriented algorithms (the LinLog in particular) should perform well according to it. Such a metric would clarify the operation performed by force-driven placement algorithms and allow evaluating them and it would have the following four features:

1. **A measure of spatialisation quality should give a score to a network layout.** The metric must measure the quality an assignment of (x,y) coordinates for the nodes of a given network spatialised with a given algorithm.
2. **We expect existing force-directed layouts to perform well according to this metric.** Since our empirical ground truth (the only one currently available) is the interpretability and usefulness of algorithms such as LinLog, we hypothesize that these algorithms accomplish a specific performance that the metric would capture and make explicit.
3. **The metric should be declined at the level of nodes and edges.** We are not only interested in evaluating spatialisations, we also want to know which nodes and edges are biased, and where we can trust the spatialisation.
4. **It should be interpretable.** We are well aware that this is a difficult goal, but the metric should tell us *what it means* to have a good spatialisation. For instance, it might be that a good visualisation is when edges are as short as possible, or that node proximity relates to the length of the shortest path, or that it represents structural equivalence. Unfortunately, as we have shown, all those explanations are wrong, so the measure should also figure out which type of information we are supposed to derive from the spatialisation of networks.

## 5. Conclusion

In this paper we have started from an empirical observation: the fact that scholars in a variety of disciplines in social and natural sciences increasingly employ network visualisations both as a preliminary way to gauge the structure of their relational datasets and as a way to convey an overview of their findings. Investigating this *evocative* power of networks, we argued that it comes from the particular way in which force-directed layouts place nodes in the space of the page or of the screen in ways that translate visually some of the most important structural features of networks.

Drawing on this intuition, and in accordance with some epistemic practices of network analysis, we described a technique which we called *Visual Network Analysis* (VNA) and which is intended to provide a step-by-step guidance in making networks readable (by choosing the right placement algorithm and tweaking its parameters) and in carrying out a systematic reading of their visual characteristics. We claimed that this approach is particularly useful for medium size network (from a few hundred to a few thousand nodes). This depends on the distinctive nature of this technique whose main advantages are (1) to highlight the topological structure of networks (rather than the connection between specific nodes); (2) to preserve some of the ambiguity inherent to relational phenomena (which is often lost in network metrics).

Finally, we set the bases for a formal validation of VNA, unfolding the difficulties related to such mathematical investigation, but also defining precisely the problem and setting the bases for its solution. With this paper we hope to help researchs to be more mindful in the use of network validations and to draw the interest of the mathematical community toward a form of analysis that is widely used, but so far insufficiently investigated.

## Appendix

### Protocol

For each network, we have generated different spatial arrangements for the nodes using the Gephi layout algorithms:

- Force Atlas 2 with default settings
- Force Atlas 2 with Lin Log activated and gravity set to 0
- Früchterman Reingold with default settings
- Random layout

Then for each network in each layout, we have computed different ways to partition the nodes in clusters:

- Modularity clustering by the Louvain method
- K-means over the spatial coordinates (x,y) of that layout, using $k$ as the number of classes found by the Louvain algorithm
- Random classes, using $k$ different classes (same number as the Louvain method)

For each pairs of partitions, we compared how similar they were by computing the Jaccard index of the set of pairs of nodes that are in the same partition. This process can be unpacked this way:

1. For a given network and a given partition of the nodes in $k$ different classes $C$
2. We build the set $S$ of all pairs of nodes *(Ni, Nj)* where the classes $Ci$ and $Cj$ are the same: $Ci = Cj$
   In other terms, this is the set of the node pairs that define the clusters.
3. To compare the two partitions *a* and *b* of the same network, we compare the sets *Sa* and *Sb* with a Jaccard index: how many pairs *(Ni, Nj)* are in common, over how many pairs are in either or both of the sets.

The Jaccard index has a value of 0 if the partitions have no node in common, and a value of 1 if they are exactly the same. Comparing the pairs of nodes has the benefit of not requiring to match each cluster of the partition *a* with a cluster of the partition *b*, which cannot always be done in a meaningful way.

### Network: CLIQUE 36

This network is a clique of 36 nodes (every node is connected to every other node).

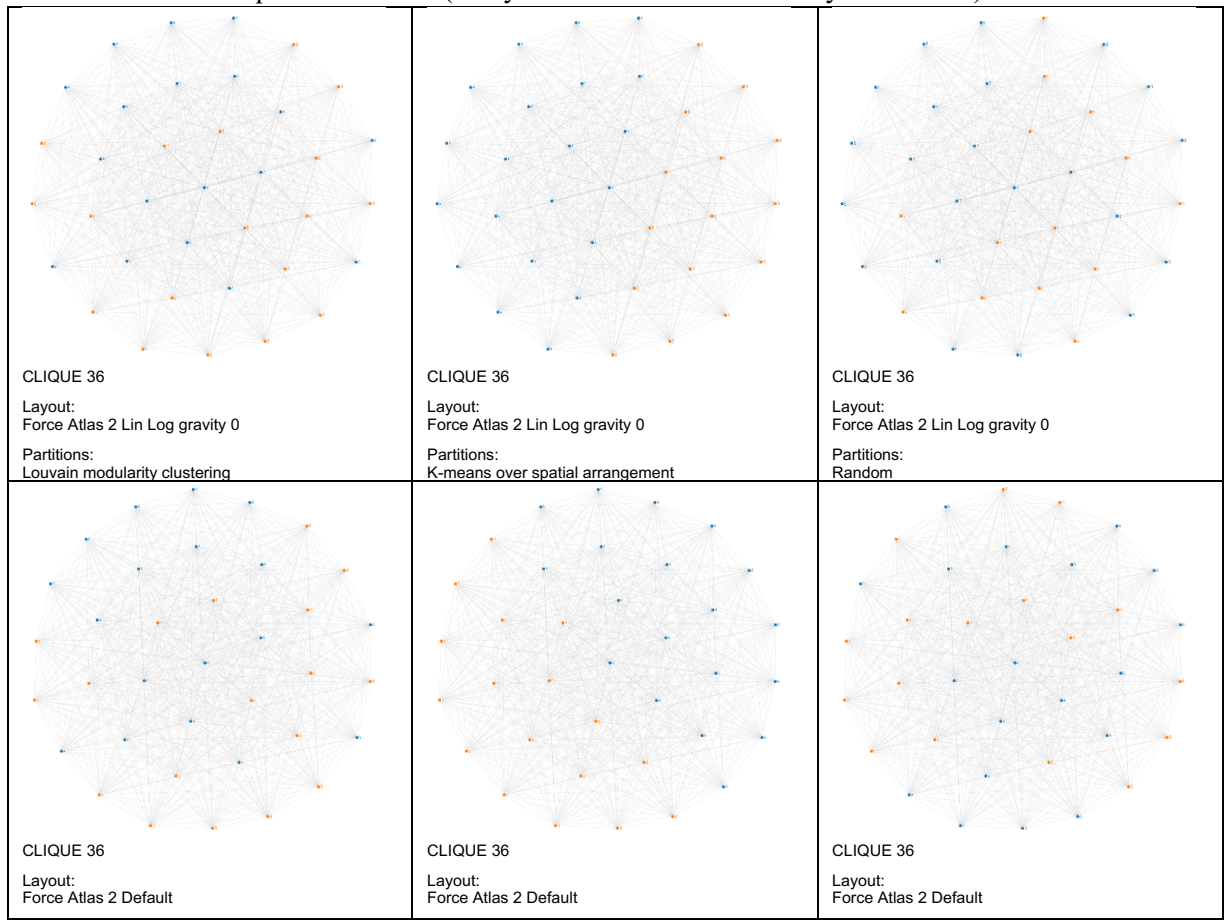

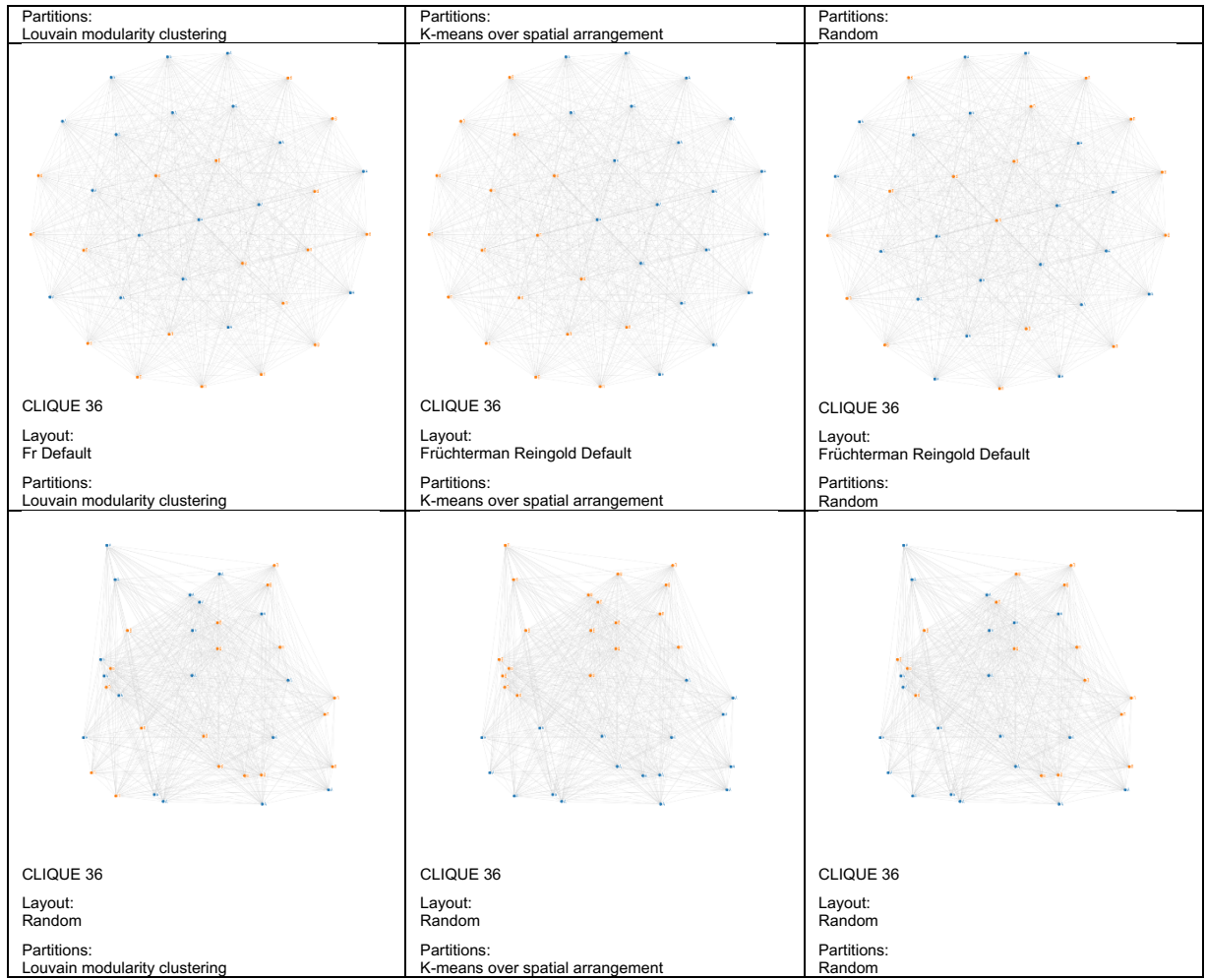

Jaccard similarities for Force Atlas 2 Lin Log gravity 0

|  | Louvain modularity | K-means (layout) | Random |
|---:|---:|---:|---:|
| Louvain modularity |  | 0.37447698744769875 | 0.3662551440329218 |
| K-means (layout) |  |  | 0.3623481781376518 |
| Random |  |  |  |

Jaccard similarities for Force Atlas 2 Default

|  | Louvain modularity | K-means (layout) | Random |
|---:|---:|---:|---:|
| Louvain modularity |  | 0.33539094650205764 | 0.33539094650205764 |
| K-means (layout) |  |  | 0.3402061855670103 |
| Random |  |  |  |

Jaccard similarities for Früchterman Reingold Default

|  | Louvain modularity | K-means (layout) | Random |
|---:|---:|---:|---:|
| Louvain modularity |  | 0.33539094650205764 | 0.34156378600823045 |
| K-means (layout) |  |  | 0.33811475409836067 |
| Random |  |  |  |

Jaccard similarities for Random layout

|  | Louvain modularity | K-means (layout) | Random |
|---:|---:|---:|---:|
| Louvain modularity |  | 0.34647302904564314 | 0.3872340425531915 |
| K-means (layout) |  |  | 0.34917355371900827 |
| Random |  |  |  |

# Network: KARATE CLUB

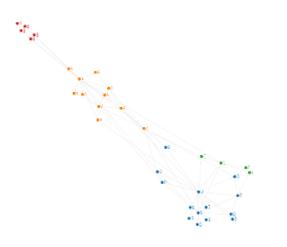

KARATE CLUB

Layout:
Force Atlas 2 Lin Log gravity 0

Partitions:
Louvain modularity clustering

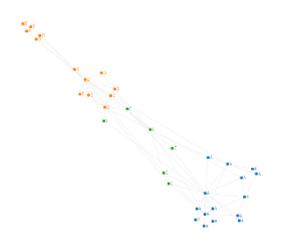

KARATE CLUB

Layout:
Force Atlas 2 Lin Log gravity 0

Partitions:
K-means over spatial arrangement

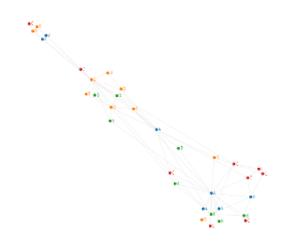

KARATE CLUB

Layout:
Force Atlas 2 Lin Log gravity 0

Partitions:
Random

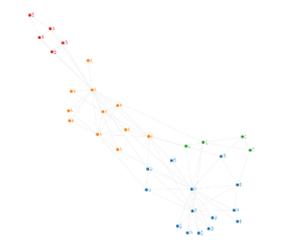

KARATE CLUB

Layout:
Force Atlas 2 Default

Partitions:
Louvain modularity clustering

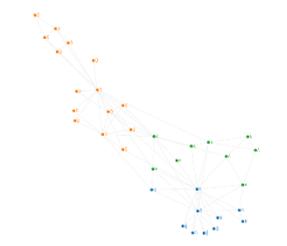

KARATE CLUB

Layout:
Force Atlas 2 Default

Partitions:
K-means over spatial arrangement

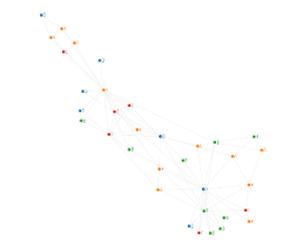

KARATE CLUB

Layout:
Force Atlas 2 Default

Partitions:
Random

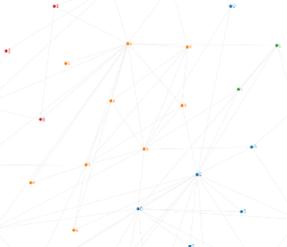

KARATE CLUB

Layout:
Fr Default

Partitions:
Louvain modularity clustering

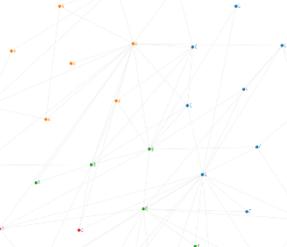

KARATE CLUB

Layout:
Früchterman Reingold Default

Partitions:
K-means over spatial arrangement

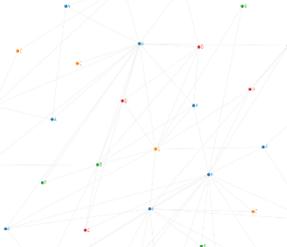

KARATE CLUB

Layout:
Früchterman Reingold Default

Partitions:
Random

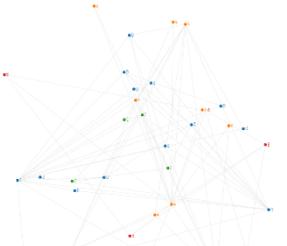

KARATE CLUB

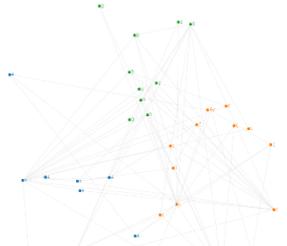

KARATE CLUB

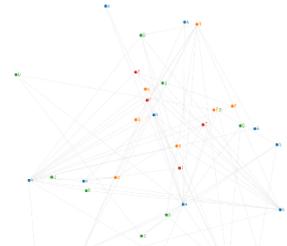

KARATE CLUB

| Layout: Random | Layout: Random | Layout: Random |
|---|---|---|
| Partitions: Louvain modularity clustering | Partitions: K-means over spatial arrangement | Partitions: Random |

Jaccard similarities for Force Atlas 2 Lin Log gravity 0

|  | Louvain modularity | K-means (layout) | Random |
|---|---|---|---|
| Louvain modularity |  | 0.4485294117647059 | 0.2354948805460751 |
| K-means (layout) |  |  | 0.23484848484848486 |
| Random |  |  |  |

Jaccard similarities for Force Atlas 2 Default

|  | Louvain modularity | K-means (layout) | Random |
|---|---|---|---|
| Louvain modularity |  | 0.5098814229249012 | 0.1971326164874552 |
| K-means (layout) |  |  | 0.20945945945945946 |
| Random |  |  |  |

Jaccard similarities for Früchterman Reingold Default

|  | Louvain modularity | K-means (layout) | Random |
|---|---|---|---|
| Louvain modularity |  | 0.3 | 0.18613138686131386 |
| K-means (layout) |  |  | 0.1821705426356589 |
| Random |  |  |  |

Jaccard similarities for Random layout

|  | Louvain modularity | K-means (layout) | Random |
|---|---|---|---|
| Louvain modularity |  | 0.22044728434504793 | 0.18900343642611683 |
| K-means (layout) |  |  | 0.2012987012987013 |
| Random |  |  |  |

# Network: JAZZ

| 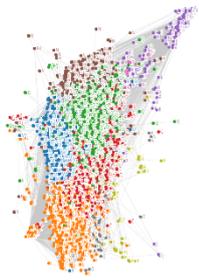 | 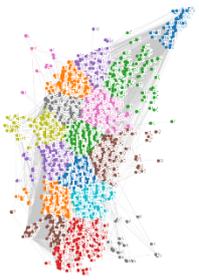 | 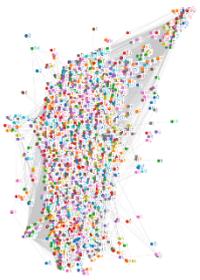 |
|---|---|---|
| JAZZ<br>Layout: Force Atlas 2 Lin Log gravity 0<br>Partitions: Louvain modularity clustering | JAZZ<br>Layout: Force Atlas 2 Lin Log gravity 0<br>Partitions: K-means over spatial arrangement | JAZZ<br>Layout: Force Atlas 2 Lin Log gravity 0<br>Partitions: Random |
| 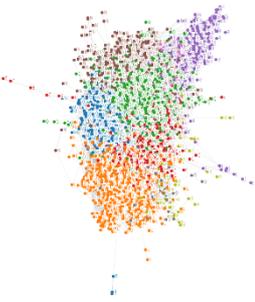 | 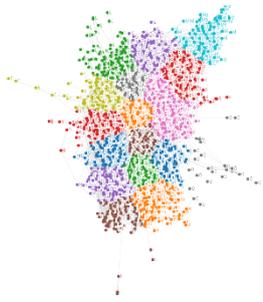 | 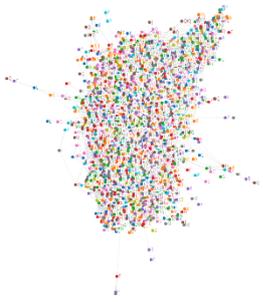 |
| JAZZ<br>Layout: Force Atlas 2 Default<br>Partitions: Louvain modularity clustering | JAZZ<br>Layout: Force Atlas 2 Default<br>Partitions: K-means over spatial arrangement | JAZZ<br>Layout: Force Atlas 2 Default<br>Partitions: Random |

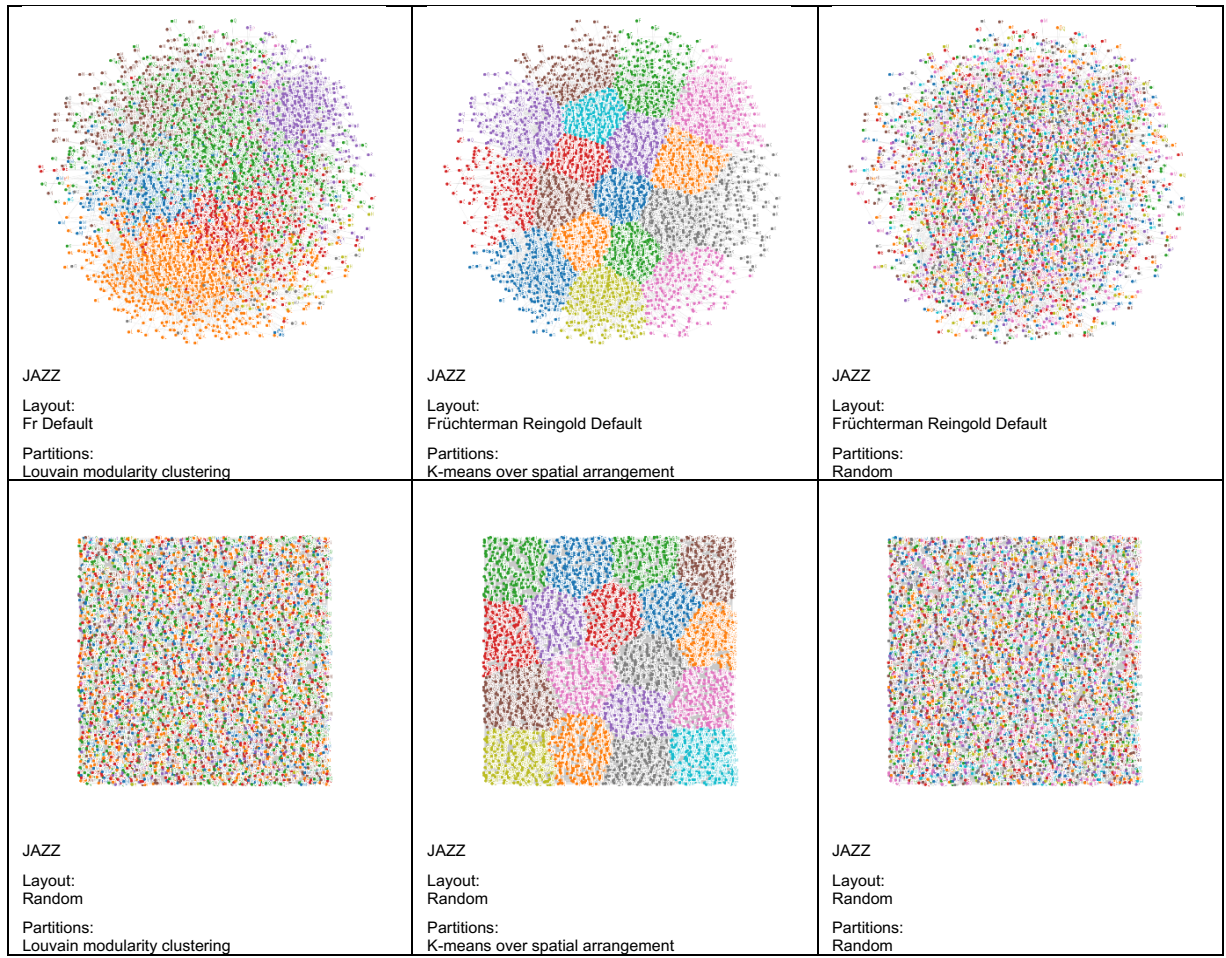

Jaccard similarities for Force Atlas 2 Lin Log gravity 0

|  | Louvain modularity | K-means (layout) | Random |
|---:|---:|---:|---:|
| Louvain modularity |  | 0.2330560156396071 | 0.04438799104420556 |
| K-means (layout) |  |  | 0.03184521017858108 |
| Random |  |  |  |

Jaccard similarities for Force Atlas 2 Default

|  | Louvain modularity | K-means (layout) | Random |
|---:|---:|---:|---:|
| Louvain modularity |  | 0.18454292538682374 | 0.044322270387411274 |
| K-means (layout) |  |  | 0.031099973208335418 |
| Random |  |  |  |

Jaccard similarities for Frŭchterman Reingold Default

|  | Louvain modularity | K-means (layout) | Random |
|---:|---:|---:|---:|
| Louvain modularity |  | 0.1550417994545389 | 0.044257343748517815 |
| K-means (layout) |  |  | 0.03167980884226725 |
| Random |  |  |  |

Jaccard similarities for Random layout

|  | Louvain modularity | K-means (layout) | Random |
|---:|---:|---:|---:|
| Louvain modularity |  | 0.04468270322090115 | 0.0302197555529323974 |
| K-means (layout) |  |  | 0.044384365553619126 |
| Random |  |  |  |

## Similarities between Louvain modularity and k-means on layout

Three remarks:

- The network CLIQUE 36 is for controlling the results. By nature, it cannot be projected on the plane in a satisfying way, and it cannot be clustered either: all nodes are strictly equivalent.
- The random layout is also presented for control. We expect that the similarity with the k-means based on it is low.
- The random partitions are for control as well. We expect that they have no meaningful similarity with either the k-means or the Louvain modularity.

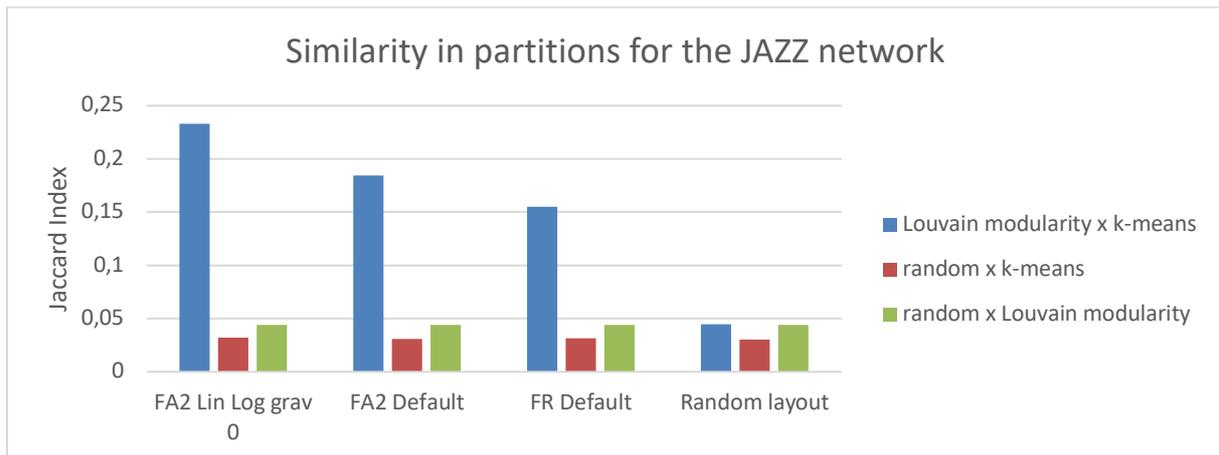

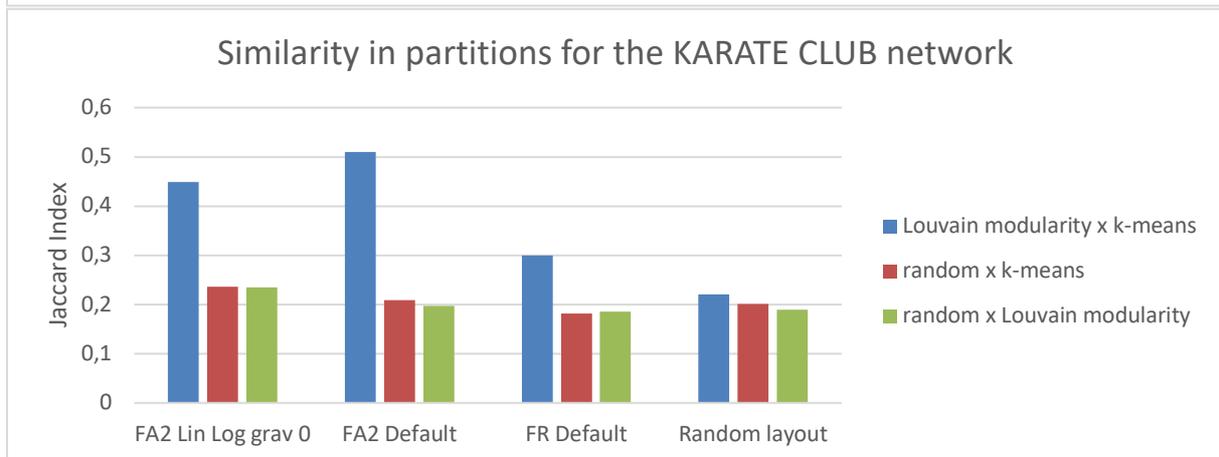

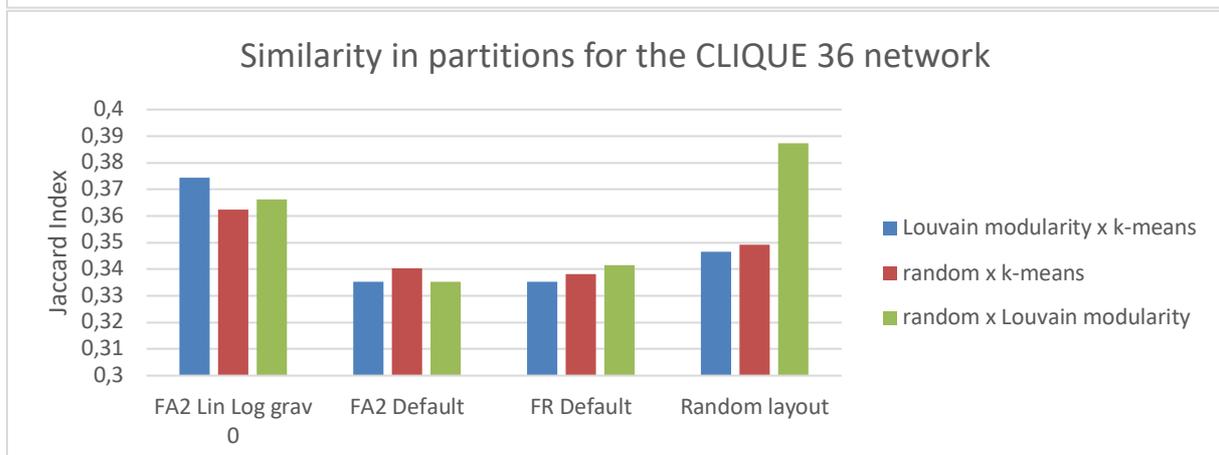